\documentclass[aps,pra,reprint,superscriptaddress]{revtex4-2}
\usepackage{graphicx}
\usepackage{amsmath}
\usepackage{xcolor}
\usepackage{braket}
 \usepackage{amssymb}
\begin{document}
\preprint{}
\title{Exact solution of a non-stationary cavity with one intermode interaction}
\author{I. Ramos-Prieto}
\affiliation{Instituto de Ciencias Físicas, Universidad Nacional Autónoma de México\\ Apdo. Postal 48-3, Cuernavaca, Morelos 62251, Mexico}
\author{R. Rom\'an-Ancheyta}
\email[e-mail:~]{ancheyta6@gmail.com}
\affiliation{Instituto Nacional de Astrofísica Óptica y Electrónica, Calle Luis Enrique Erro No. 1, Santa María Tonantzintla, Pue., 72840, Mexico}
\author{J. Récamier}
\affiliation{Instituto de Ciencias Físicas, Universidad Nacional Autónoma de México\\ Apdo. Postal 48-3, Cuernavaca, Morelos 62251, Mexico}
\author{H. M. Moya-Cessa}
\affiliation{Instituto Nacional de Astrofísica Óptica y Electrónica, Calle Luis Enrique Erro No. 1, Santa María Tonantzintla, Pue., 72840, Mexico}
\date{\today}
\begin{abstract}
A non-stationary one-dimensional cavity can be described by the time-dependent and multi-mode effective Hamiltonian of the so-called dynamical Casimir effect. Due to the non-adiabatic boundary conditions imposed in one of the cavity mirrors, this effect predicts the generation of real photons out of vacuum fluctuations of the electromagnetic field. Such photon generation strongly depends on the number of modes in the cavity and their intermode couplings. Here, by using an algebraic approach, we show that for any set of functions parameterizing the effective Hamiltonian,  the corresponding time-dependent Schr\"odinger equation admits an exact solution when the cavity has one intermode interaction. With the exact time evolution operator, written as a product of eleven exponentials, we obtain the average photon number in each mode, a few relevant observables and some statistical properties for the evolved vacuum state.
\end{abstract}
\maketitle
\section{Introduction}
  With a simple but nontrivial model of a one-dimensional cavity of variable length, in 1970, Gerald~T.~Moore showed the possibility of generating real photons from the electromagnetic field's vacuum state~\cite{Moore_1970}.
    This phenomenon is known today as the dynamical, or nonstationary, Casimir effect (DCE), names coined in~\cite{Yablonovith_1989,Schwinger_1992} and~\cite{Dodonov_1989} because it is directly related to the vacuum fluctuations~\cite{Nation_2012}. Since then, the DCE has been associated with other exotic phenomena~\cite{Berdiyorov_2014,Dodonov_2020_a}
    like quantum particle emission from black holes~\cite{Fulling_1976,Davis_1977}.
    In addition to the experimental observation of the DCE in superconducting quantum circuits \cite{Wilson_2011,Pasi_2013},
    analogous systems~\cite{Fujii_2011,Roman_2017} have also been proposed to simulate the DCE and study, for example, the impact of optical nonlinearities~\cite{Roman-Ancheyta:17} and energy losses~\cite{Roman_2018} upon the vacuum photon generation.
    During the last fifty years, the conceptual and technical richness of the DCE has generated endless proposals~\cite{Dodonov_2010} and questions~\cite{Milton_2004} from multiple perspectives that continue to revive some aspects not yet explored \cite{Dodonov_2020_b}.

   From the theoretical point of view, numerical approaches to solve the DCE quickly reveal the complexity of obtaining the photon production rate~\cite{Ruser_2006,Ruser_2006_a}. This complexity increases when considering contributions from the unavoidable intermodal interaction, especially in one-dimensional cavities~\cite{Li_2002}. In such scenario, it is possible to determine some observables of the system and compare them with analytical approximations~\cite{Dodonov_1996,Dodonov_1998,Dodonov_Analytical_Solutions}. However, the lack of knowledge of an exact time evolution operator reveals some limitations on these types of numerical solutions.

    In this work, by using the Wei-Norman approach~\cite{Wei-Norman}, we present an exact and general solution of the DCE when the non-stationary cavity has two coupled modes. Of particular importance, we study the photon generation when only one of these modes is on parametric resonant conditions. We show how it is possible to factorize, or disentangle, the total system's evolution operator for any set of functions that parameterize the corresponding time-dependent effective Hamiltonian.

    We structured the paper as follows. In Sec.~\ref{Theory} we show that the set of operators of the corresponding time-dependent Hamiltonian, resembling the Hamiltonian of two harmonic oscillators with arbitrary interactions \cite{Urzua_2019}, is closed under commutation, that allows us to write the system's evolution operator in terms of functions that satisfy a finite number of nonlinear coupled differential equations. In Sec.~\ref{section_3}, we evaluate the average number of photons in each mode, the dispersion of the quadratures and discuss the validity of our results by comparing them with a purely numerical calculation done using QuTiP \cite{Johansson_2012}. Due to the complexity and number of coupled nonlinear differential equations, we also tested our results by means of the quantum universal invariant approach \cite{Dodonov_2000}. In Sec.~\ref{conclusions} we present our conclusions. Separated from the main text, we show, in two appendices the entire commutation relations and the corresponding system of differential equations.

\section{Wei-Norman approach for the DCE}\label{Theory}
Usually, the DCE is  studied using the effective Hamiltonian approach~\cite{Plunien_PRA_1998,Law_1994}. There, the Hamiltonian for an  electromagnetic field in a cavity, having $N$ modes and a moving mirror, is given by~\cite{Law_1994}:
\begin{equation}\label{eq:heff}
\begin{split}
\hat{H}_{\rm eff}(t)&=\sum_{k=1}^{N} \omega_k(t)\hat a_k^{\dagger}\hat a_k +\frac{\mathrm{i}}{4}\frac{\dot{q}(t)}{q(t)}\sum_{k=1}^{N} (\hat a_k^{\dagger2}-\hat a_k^{2})\\
&+\frac{\mathrm{i}}{2}\frac{\dot{q}(t)}{q(t)}\sum_{
j,k \atop j\neq k}^{N} \mu_{j,k}\left(\hat a_k^{\dagger}\hat a_j^{\dagger}+ \hat a_k^{\dagger}\hat a_j - \hat a_k\hat a_j^{\dagger}-\hat a_k \hat a_j\right).
\end{split}
\end{equation}
For each mode $k$, $\hat{a}_k$, $\hat{a}_k^\dagger$ are the usual bosonic annihilation and creation operators in the Schr\"odinger picture satisfying  $[\hat{a}_k,\hat{a}_j^\dagger]=\delta_{k,j}$. The time-dependent real function $q(t)$ represents the trajectory of one of the cavity mirrors, and $\dot{q}(t)={\rm d}{q(t)}/{\rm d}{t}$ it's time-derivative.
The first term on the right-hand side of $\hat{H}_{\rm eff}(t)$ denotes a set of quantum harmonic oscillators with instantaneous cavity frequency $\omega_k(t)={k\pi}/{q(t)}$ \cite{Law_1994}. The second term is known as the squeezing Hamiltonian, and the last one represents all the intermode interactions where,
\begin{equation}
\mu_{j,k}=(-1)^{j+k}\frac{kj}{j^2-k^2}\left(\frac{k}{j}\right)^{1/2}.
\end{equation}
It is well known that for the single-mode case $(N=1)$, the non-stationary cavity has an exact~\cite{Dodonov_Analytical_Solutions,Dodonov_2010} and sometimes explicit analytic~\cite{Dodonov_2013_PhyScr,Roman_2017,Roman_2018} solution, where the field evolves into a squeezed vacuum state with an exponential photon growth. Here, we are particularly interested in a situation where the non-stationary cavity has two modes ($N=2$)~\cite{AV_Dodonov_2001_PLA,AV_Dodonov_2020_PLA} so that the above Hamiltonian can be rewritten in normal order as:
\begin{equation}\label{eq:hwff2m}
\hat H_{\rm eff}(t)= \sum_{n=1}^{11} f_n(t) \hat X_n,
\end{equation}
with the functions $f_n(t)$ and the time-independent operators $\hat{X}_n$ given as:
\begin{center}
\begin{equation}\label{eqX}
\begin{tabular}{l l}
$f_1(t) = \mathrm{i}\frac{\dot{q}(t)}{4q(t)}$, & $ \hat X_1=\hat a_1^{\dagger 2}$, \\
$f_2(t) = \mathrm{i}\frac{\dot{q}(t)}{4q(t)}$, & $ \hat X_2 = \hat a_2^{\dagger 2}$, \\
$ f_3(t) =\frac{\mathrm{i}}{2}\left(\mu_{1,2}+\mu_{2,1}\right)\frac{\dot{q}(t)}{q(t)}$, & $ \hat X_3 = \hat a_1^{\dagger} \hat a_2^{\dagger} $, \\
$ f_4(t) = -\frac{\mathrm{i}}{2}\left(\mu_{1,2}-\mu_{2,1}\right)\frac{\dot{q}(t)}{q(t)}$, & $ \hat X_4 = \hat a_1^{\dagger}\hat a_2$, \\
$ f_5(t) = -f_4(t)$, & $ \hat X_5 = \hat a_1 \hat a_2^{\dagger} $, \\
$f_6(t) = \frac{\pi}{q(t)} $, & $ \hat X_6 = \hat a_1^{\dagger}\hat a_1 $, \\
$f_7(t) = \frac{2\pi}{q(t)} $, & $  \hat X_7 = \hat a_2^{\dagger}\hat a_2 $, \\
$ f_8(t) = -f_1(t)$, & $ \hat X_8 = \hat a_1^2 $, \\
$ f_9(t) =  -f_2(t)$, & $ \hat X_9 = \hat a_2^2 $, \\
$ f_{10}(t) = - f_3(t)$, & $ \hat X_{10} = \hat a_1 \hat a_2 $, \\
$ f_{11}(t) = 0 $, & $ \hat X_{11} = \hat{\mathbb{I}}$. \\
\end{tabular}
\end{equation}
\end{center}
$\hat{\mathbb{I}}$ is the identity operator. Obviously, $\hat{H}_{\rm eff}(t)$ does not commute with itself at different times, i.e.,
\begin{equation}
\big[\hat{H}_{\rm eff}(t_i),\hat{H}_{\rm eff}(t_j)\big]\neq 0, \quad \forall \quad t_i,\, t_j.
\end{equation}
This means that standard textbook methods~\cite{sakurai_napolitano_2017} to obtain the corresponding time evolution operator can not be efficiently implemented, and one might think of adopting  approximate solutions. Fortunately, it is relatively easy but cumbersome to show that the set of operators $\hat{X}_n$ in Eq.~\eqref{eq:hwff2m} is closed under commutation (see appendix \ref{Apen_A}). Therefore, we can invoke the Wei-Norman theorem and write, without any approximation, the exact time evolution operator $\hat{U}(t)$ as a product of exponentials~\cite{Wei-Norman}
 \begin{equation}\label{eq:tevol}
 \hat U(t)= \prod_{j=1}^{11} \exp\left[\alpha_j(t) \hat X_j\right].
 \end{equation}
  The task now is to find the set of ordinary differential equations satisfied by the complex, time-dependent functions $\alpha_j(t)$.
 In order to do that, we first substitute \eqref{eq:tevol} in the Schr\"odinger equation ${\rm i}\partial_t\hat{U}(t)=\hat{H}_{\rm eff}(t)\hat{U}(t)$ and after some elementary algebra, we obtain:
\begin{subequations}
\begin{eqnarray}
\mathrm{i}\frac{\partial \hat{U}(t)}{\partial t} & = & {\rm i} \bigg[ \dot \alpha_1(t) \hat{X}_1+ \dot \alpha_2(t) \hat{X}_2 + \dot\alpha_3(t) \hat{X}_3\nonumber\\&&+\sum_{j=4}^{11}\dot{\alpha}_j(t)\prod_{n=1}^{j-1}e^{\alpha_n(t)\hat{X}_n}\hat{X}_{j}\prod_{n=1}^{j-1}e^{-\alpha_n(t)\hat{X}_n}\bigg] \hat U(t),\nonumber \\ & = &\mathrm{i}\bigg[\sum_{j=1}^{11}\sum_{n=1}^{11} \dot \alpha_j(t)M_{nj} \hat X_n \bigg] \hat{U}(t),\label{dUa}\\
&=& \bigg[\sum_{n=1}^{11} f_n(t) \hat X_n\bigg]\hat{U}(t).\label{dUb}
  \end{eqnarray}
\end{subequations}
Since the commutators $[\hat{X}_1,\hat{X}_2]=[\hat{X}_1,\hat{X}_3]=[\hat{X}_2,\hat{X}_3]=0$, the first three terms of the above equation are immediate; however, this is not the case for $j\geq4$. After a bit more of algebra of operators, it is more or less easy to establish the differential equations dictated by Eqs.~(\ref{dUa}) and (\ref{dUb}). For instance, $n=1$ in Eq.~(\ref{dUa}) implies $\hat X_1=\hat a_1^{\dagger 2}$, then the expressions for $M_{1j}$ are:
\begin{equation}\label{M_1j}
\begin{split}
M_{11}=&1,\\
M_{12}=&0,\\
M_{13}=&0,\\
M_{14}=&-\alpha_3,\\
M_{15}=&\alpha _3 \alpha _4^2-2 \alpha _1 \alpha _4,\\
M_{16}=&\alpha _3 \alpha _5 \alpha _4^2+\alpha _3 \alpha _4-2 \alpha _1 \alpha _5 \alpha _4-2
   \alpha _1,\\
M_{17}=&-\alpha _3 \alpha _5 \alpha _4^2-\alpha _3 \alpha _4+2 \alpha _1 \alpha _5 \alpha _4,\\
M_{18}=&e^{-2 \alpha _6}\big(4 \alpha _1^2-4\alpha _3 \alpha _4 \alpha _1+\alpha _3^2 \alpha _4^2\big),\\
M_{19}=&e^{-2 \alpha _7}\big(\alpha _3^2+\alpha _4^2 \alpha _5^2 \alpha _3^2+2\alpha _4 \alpha _5 \alpha _3^2-4\alpha _1 \alpha _4 \alpha _5^2 \alpha _3\\&-4\alpha _1 \alpha _5 \alpha _3+4\alpha _1^2 \alpha_5^2\big),\\
M_{1\ 10}=&e^{-(\alpha _6+\alpha _7)}\big(2\alpha _3 \alpha _1-4\alpha _5 \alpha _1^2-\alpha _3^2 \alpha _4-\alpha _3^2 \alpha _4^2 \alpha _5\\&+4\alpha _3 \alpha _4 \alpha _5 \alpha_1\big),\\
M_{1\ 11}=&0.
\end{split}
\end{equation}
As we go along with the index $n$, the number of terms and their handling gets more involved, see appendix~\ref{Apen_B} for the explicit form of the remaining $M_{nj}$. Nevertheless, by using Eq.~\eqref{dUb} we can write the system of differential equations in a matrix form, and taking the inverse of such matrix we get the corresponding equations of motion for the functions $\alpha_j(t)$:
\begin{equation}\label{alphas}
\dot{\bm{\alpha}}(t)+\mathrm{i}\bm{M}^{-1}\bm{f}(t)=0,
\end{equation}
where $\dot{\bm{\alpha}}(t)=\left(\dot{\alpha}_1(t),\dot{\alpha}_2(t),\dots,\dot{\alpha}_{11}(t)\right)^{T}$ and $\bm{f}(t)=\left(f_1(t),f_2(t),\dots,f_{11}(t)\right)^{T}$.  Here we would like to mention that the evolution operator $\hat{U}(t)$ in Eq.~\eqref{eq:tevol} is by definition unitary; however, it depends on whether or not the inverse of $\bm{M}$, denoted by $\bm{M}^{-1}$, exists. Recall that this matrix $\bm{M}$ depends on the $\alpha_j(t)$ coefficients and that the system of coupled differential equations is highly nonlinear, see Eq.~\eqref{Eq:alphasM}.

Once we have solved by either analytical or numerical means the differential equations for the functions $\alpha_j(t)$, we have an exact expression for the time evolution operator given in Eq.~\eqref{eq:tevol}. This is a significant result that can help our understanding of the DCE with two coupled modes because it solves the problem for any trajectory $q(t)$. Furthermore, when the non-stationary cavity has more than two coupled modes, we speculate that the procedure followed up to this point can be, in principle, generalized since the operators' set of \eqref{eq:heff} seems to close under commutation. In fact, we tested such hypothesis for up to $N=3$. This observation comes from the fact that the quadratic structure of $\hat{H}_{\rm eff}(t)$ has only time-dependent bilinear interactions.

Before concluding this section, we obtain the explicit expressions of each mode's creation and annihilation operators in the Heisenberg picture. This is necessary because, in the next section, we will compute the average photon number from the vacuum state for each mode. Using the evolution operator given in Eq.~\eqref{eq:tevol}, and Eq.~\eqref{alphas} we get the creation-annihilation operators in the Heisenberg picture as:
\begin{equation}\label{eq_tj}
\begin{split}
\hat{a}_1(t) &= t_{11}(t)\hat{a}_1 + t_{12}(t)\hat{a}_1^{\dagger}+t_{13}(t)\hat{a}_2+t_{14}(t)\hat{a}_2^{\dagger},\\
\hat{a}_1^{\dagger}(t) &= t_{21}(t)\hat{a}_1 +t_{22}(t)\hat{a}_1^{\dagger}+t_{23}(t)\hat{a}_2+t_{24}(t)\hat{a}_2^{\dagger},\\
\hat{a}_2(t) &= t_{31}(t)\hat{a}_1+t_{32}(t)\hat{a}_1^{\dagger}+t_{33}(t)\hat{a}_2+t_{34}(t)\hat{a}_2^{\dagger},\\
\hat{a}_2^{\dagger}(t) &= t_{41}(t)\hat{a}_1+t_{42}(t)\hat{a}_1^{\dagger}+t_{43}(t)\hat{a}_2+t_{44}(t)\hat{a}_2^{\dagger},
\end{split}
\end{equation}
where the time-dependent coefficients $t_{ij}(t)$ are functions of $\alpha_j(t)$, their explicit form is given in appendix~\ref{Apen_B}. Due to the unitarity of the evolution operator $\hat{U}(t)$, the following relations must hold:
\begin{equation}
\begin{split}
t_{11}&= t_{22}^{*}, \ \ \ t_{12}= t_{21}^{*}, \ \ \ t_{13}= t_{24}^{*}, \ \ \ t_{14}= t_{23}^{*},\\
t_{31}&= t_{42}^{*},\ \ \ t_{32}= t_{41}^{*}, \ \ \ t_{33}= t_{44}^{*}, \ \ \ t_{34}= t_{43}^{*},
\end{split}
\end{equation}
where we omit the explicit time dependence to simplify the notation. We have also ensured that these relationships are satisfied at any time by a numerical evaluation~\cite{Johansson_2012}.

So far, we have not specified a particular law of motion for $q(t)$, and we have not used the explicit form of the functions $f_n(t)$. This means that as long as the inverse matrix in \eqref{alphas} exists, our algebraic method should work for whatever set of time-dependent functions parameterizing the Hamiltonian of \eqref{eq:heff}.

\section{Temporal evolution}\label{section_3}
This section evaluates the average number of photons from the vacuum state for each mode, the corresponding quadratures variance, and the Mandel parameter. It is usually convenient to choose the mirror's trajectory as~\cite{Law_1994}:
\begin{equation}
q(t)=L\exp\left[\frac{q_0}{L} \sin\left(\omega_dt+\phi\right)\right],
\end{equation}
where $L$ is the cavity length when $t=\phi=0$. Note that, in the weak perturbation regime ($q_0\ll L$) the path of the mirror approaches a simple harmonic motion with modulation frequency $\omega_d$ and initial phase $\phi$. We want to emphasize that there are analytical solutions to this system in the literature, but only under particular situations and approximations. For instance, when in and out of resonance conditions are considered. However, an exact solution in the most general case of two coupled modes, like the one we present in the previous section, seems non-existent, see \cite{AV_Dodonov_2001_PLA} and references within \cite{Dodonov_2020_b}.

To see if there are squeezing effects present in the system we define the field quadratures for each mode as~\cite{gerry_knight_2004}
\begin{equation}
\begin{split}
\hat{Q}_1= ( \hat a_1 +\hat a_1^{\dagger})/\sqrt{2}&,\quad \hat{P}_1 = {i}(\hat a_1^{\dagger}-\hat a_1)/\sqrt{2},\\
\hat{Q}_2= ( \hat a_2 +\hat a_2^{\dagger})/\sqrt{2}&,\quad \hat{P}_2 = {i}(\hat a_2^{\dagger}-\hat a_2)/\sqrt{2}.
\end{split}
\end{equation}
Hence, using the evolution operator given by Eq.~\eqref{eq:tevol} we obtain the quadratures in the Heisenberg representation as:
\begin{subequations}
\begin{eqnarray}
\hat Q_1(t) &=&\big[(t_{11}+t_{21})\hat a_1 + (t_{12}+t_{22})\hat a_1^{\dagger}\nonumber\\ && \qquad+(t_{13}+t_{23})\hat a_2 +(t_{14}+t_{24})\hat a_2^{\dagger}\big]/{\sqrt{2}},\qquad\\
\hat P_1(t) &=&i\big[(t_{21}-t_{11})\hat a_1 +(t_{22}-t_{12})\hat a_1^{\dagger}\nonumber\\ &&\qquad+(t_{23}-t_{13})\hat a_2 +(t_{24}-t_{14})\hat a_2^{\dagger}\big]/\sqrt{2},\qquad\\
\hat Q_2(t) &=&\big[(t_{31}+t_{41})\hat a_1 + (t_{32}+t_{42})\hat a_1^{\dagger}\nonumber\\ &&\qquad+(t_{33}+t_{43})\hat a_2 +(t_{34}+t_{44})\hat a_2^{\dagger}\big]/\sqrt{2},\qquad\\
\hat P_2(t) &=& {i}\big[(t_{41}-t_{31})\hat a_1 +(t_{42}-t_{32})\hat a_1^{\dagger}\nonumber\\ &&\qquad+(t_{43}-t_{33})\hat a_2 +(t_{44}-t_{34})\hat a_2^{\dagger}\big]/\sqrt{2},\qquad
\end{eqnarray}
\end{subequations}
where the coefficients $t_{ij}$ are given in \eqref{coeff_tjk} of the  appendix~\ref{Apen_B}. For $\hat Q_1^2(t)$, $\hat P_1^2(t)$, $\hat Q_2^2(t)$ and $\hat P_2^2(t)$ we get more cumbersome expressions that we do not write. To unambiguously  discuss the dynamical Casimir effect, it is necessary to take as the initial state the two-modes' vacuum state $|{0,0}\rangle\equiv|{0}\rangle_1\otimes|{0}\rangle_2$.
The variance of the quadratures calculated with respect to this initial state reduces to:
\begin{subequations}
\begin{eqnarray}
(\Delta Q_1(t))^2  &=& [ (t_{11}+t_{21})(t_{12}+t_{22})+\nonumber\\ &&\qquad\qquad(t_{13}+t_{23})(t_{14}+t_{24})]/2,\quad\\
(\Delta P_1(t))^2  &=& [(t_{11}-t_{21})(t_{22}-t_{12})+\nonumber\\
&&\qquad\qquad(t_{13}-t_{23})(t_{24}-t_{14})]/2,\quad\\
(\Delta Q_2(t))^2  &=& [(t_{31}+t_{41})(t_{32}+t_{42})+\nonumber\\
&&\qquad\qquad(t_{33}+t_{43})(t_{34}+t_{44})]/2,\quad\\
(\Delta P_2(t))^2  &=&[(t_{31}-t_{41})(t_{42}-t_{32})+\nonumber\\
&&\qquad\qquad(t_{33}-t_{43})(t_{44}-t_{34})]/2,
\end{eqnarray}
\end{subequations}
where $(\Delta O)^2\equiv \langle  \hat O^2\rangle-\langle \hat O\rangle^2$.

\begin{figure}[t!]
\begin{center}
\includegraphics[width=\linewidth]{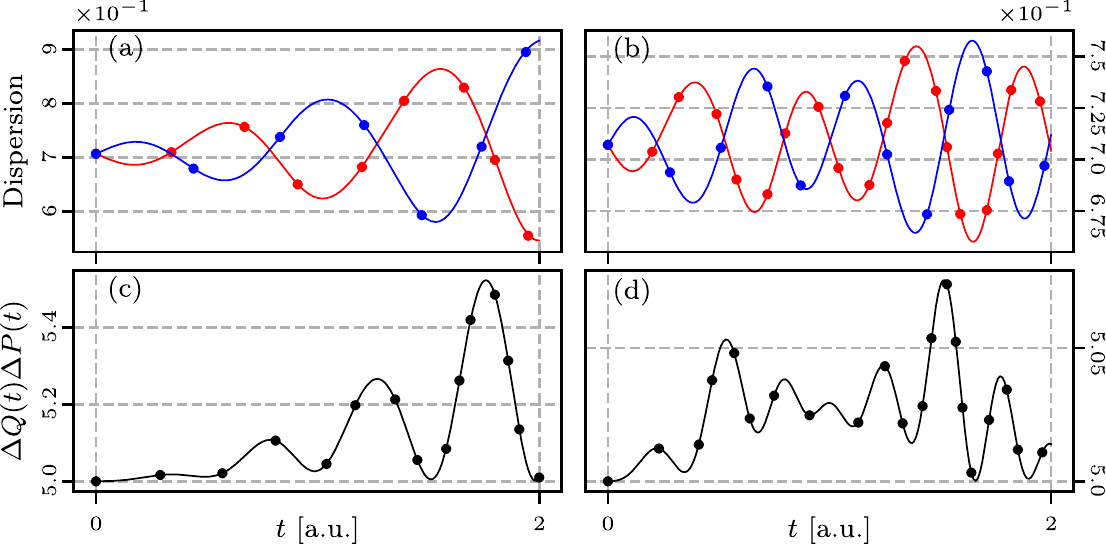}
\caption{Dispersion $\Delta Q_j(t)$ (blue) and $\Delta P_j(t)$ (red), for $j=1$ (a) and $j=2$ (b), with respect to the initial vacuum state. Product of the dispersions $\Delta Q_1(t) \Delta P_1(t)$ (c) and $\Delta Q_2(t) \Delta P_2(t)$ (d).  System's parameters are $L=1$, $q_0={1}/{12}$, $\phi=0$, and $\omega_d=2\pi$. In all cases, solid lines (solid dots) are the exact (purely numeric) results. Numerical simulations were performed in QuTiP \cite{Johansson_2012}.}
\label{Fig_1}
	\end{center}
\end{figure}

In Figs.~\ref{Fig_1}-(a) and~\ref{Fig_1}-(b) we show, respectively, the temporal evolution of $\Delta Q_1(t)$, $\Delta P_1(t)$ and $\Delta Q_2(t)$, $\Delta P_2(t)$. These dispersions oscillate around the lower bound $1/\sqrt{2}\approx 0.7$ of the corresponding Glauber coherent states, which is also the same for the ground state of the cavity at the initial time, i.e., $\Delta Q_j(0) = \Delta P_j(0) = {1}/{\sqrt{2}}$. As time passes squeezing effects start to appear, alternating between each mode's quadratures due to the rotating behavior of the state in the quatum optical phase-space~\cite{Schleich2001}. For the fundamental mode, which is under resonance conditions,  the amplitude of the oscillations increases with time, in contrast, for the second mode (or mode number two) which is not under resonance conditions, we see similar oscillations with smaller amplitude around the coherent state limit.
In Fig.~\ref{Fig_1}-(c) and Fig.~\ref{Fig_1}-(d), the product $\Delta Q_j(t)\Delta P_j(t)$ of the uncertainties is always larger than or equal to ${1}/{2}$ as it should be. There are also some instants of time where such product attains its minimum possible value for both modes.
It is important to mention that the set of parameters used in Fig.~\ref{Fig_1} is such that the fundamental mode with unperturbed frequency $\omega_1=\pi$ is under parametric resonance, i.e., the frequency modulation is $\omega_d=2\pi$. Evidently, the second mode has an unperturbed frequency $\omega_2=2\pi$, which is not in parametric resonant conditions with $\omega_d$. In the following, we will see that these conditions permit a large photon generation in, at least, the fundamental mode of the cavity field.

Now, we will calculate the average photon number $\langle \hat n_j(t)\rangle$ for each mode $j$. The number operator $\hat n_j\equiv \hat{a}_j^\dagger\hat{a}_j$ in the Heisenberg picture is:
\begin{eqnarray}
\hat{n}_j(t) = \hat{U}^{\dagger}(t)\hat n_j\hat{U}(t) = \hat{U}^{\dagger}(t)\hat{a}_j^{\dagger}\hat{a}_j\hat{U}(t) = \hat{a}_j^{\dagger}(t)\hat{a}_j(t),\qquad\,
\end{eqnarray}
where $\hat{U}(t)$ is given in \eqref{eq:tevol}, and $\hat{a}_j(t)$, $\hat{a}_j^{\dagger} (t)$ can be found in \eqref{eq_tj}. Taking the expectation value with respect to the system's initial state yields,
\begin{equation}\label{exact_n1_n2}
\langle\hat{n}_1(t)\rangle_0 = |t_{12}|^2+|t_{14}|^2,
\,
\langle\hat{n}_2(t)\rangle_0 = |t_{32}|^2+|t_{34}|^2,
\end{equation}
where $\langle \hat{O}\rangle_0\equiv\langle 0,0|\hat{O}|0,0\rangle$.

\begin{figure}[t!]
\begin{center}
\includegraphics[width=\linewidth]{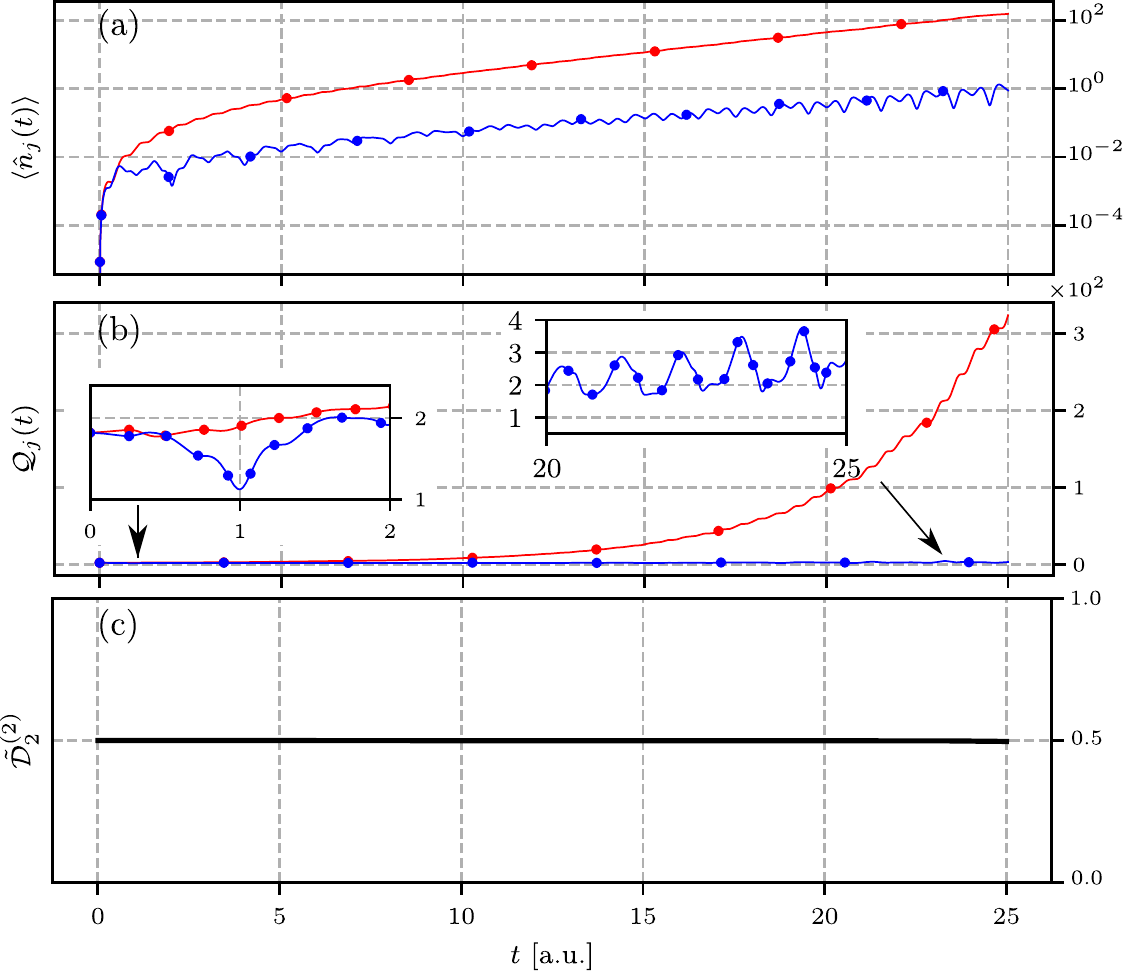}
\caption{(a) Average value of the number operator $\langle\hat{n}_1(t)\rangle$ (red) and $\langle\hat{n}_2(t)\rangle$ (blue) using Python (dots) and the semi-analytical results (lines). (b) Mandel $\mathcal{Q}$ parameter for mode one (red) and mode two (blue) with $\omega_d=2\pi$, on resonance condition for photon generation of mode one. Numerical results obtained with Python (dots) and semi-analytic results (solid lines). The Hamiltonian parameters are: $L=1$, and $q_0={1}/{12}$. (c) Quantum universal invariant (see~\eqref{ui}) as a function of time; same parameters as in (a) and (b).}
\label{Fig_2}
\end{center}
\end{figure}

Figure~\ref{Fig_2}-(a) shows the temporal evolution of the abovementioned average number of photons obtained from a converged purely numerical solution using Python~\cite{Johansson_2012}(dots) and the exact results of \eqref{exact_n1_n2} following a semi-analytic approach (continous line), i.e., we use the exact expression for the evolution operator, but we have solved the differential equations for the functions $\alpha_j(t)$ with the program Mathematica~\cite{Mathematica}. We see an excellent agreement between both approaches. Since the modulating frequency $\omega_d$ is in parametric resonance with the fundamental mode, $\omega_d=2\omega_1$, the average number of photons for this mode increases rapidly while the average number of photons for the second mode increases, still exponentially, but much more slowly.

To evaluate the Mandel $\mathcal{Q}$ parameter~\cite{Mandel_1979} for mode $j$ we use
\begin{equation}
\mathcal{Q}_j = \big({\langle \hat n_j^2(t)\rangle -\langle \hat n_j(t)\rangle^2}\big)/{\langle \hat n_j(t)\rangle}.
\end{equation}
A state with $\mathcal{Q}$ in the range $0\leq\mathcal{Q}<1$ has sub-Poissonian statistics (quantum behavior),  and for $\mathcal{Q}>1$ super-Poissonian (classical behavior). For a coherent state $\mathcal{Q}=1$. Taking the expectation values by using an initial state given by the vacuum $ {|0,0\rangle}$ we get
\begin{subequations}
\begin{eqnarray}
\langle\hat{n}_1^2(t)\rangle_0 &=& ( |t_{12}|^2+|t_{14}|^2)^2 +|t_{11}t_{23}+t_{13}t_{21}|^2\nonumber\\ && +2(|t_{11}|^2 |t_{12}|^2+|t_{14}|^2 |t_{13}|^2),\qquad \\
\langle\hat{n}_2^2(t)\rangle_0 &=& (|t_{41}|^2+|t_{43}|^2)^2+|t_{31}t_{43}+t_{33}t_{41}|^2 \nonumber\\ && +2(|t_{41}|^2 |t_{42}|^2+|t_{43}|^2 |t_{33}|^2).\quad
\end{eqnarray}
\end{subequations}

In Fig.~\ref{Fig_2}-(b) we show the Mandel $\mathcal{Q}$ parameter for mode one (red) and mode two (blue) using Python (dots) and the analytical result (solid lines) as a function of time with an initial state $|0\rangle \otimes |0\rangle$.  At the beginning of the evolution the photon number is zero in both modes, the Mandel $\mathcal{Q}$ parameter is the same for both modes and it is near two (see the inset in the figure). As time evolves the average photon number for mode one increases rapidly (see Fig.~\ref{Fig_2}-(a)) and the corresponding Mandel $\mathcal{Q}$ parameter also increases, after about 20 time units,  the fundamental mode has reached about $10^2$ photons and it's Mandel parameter shows a super Poissonian statistics. On the other hand, the average number of photons for mode two  remains of the order of one and the corresponding Mandel $\mathcal{Q}$ parameter remains near the value of a coherent state.

Finally, although the Hamiltonian described here is explicitly time-dependent, there may be quantum universal invariants provided the Hamiltonian is a quadratic form of position and moment operators~\cite{Dodonov_2000}. In particular, for the case of two degrees of freedom and bosonic operators, an invariant associated with Hamiltonian in \eqref{eq:heff} can be written as~\cite{Dodonov_2000}:
\begin{equation}\label{ui}
    \tilde{\mathcal{D}}_{2}^{(2)}=\sum_{j=1}^2\big(\sigma_{N_j}+\frac{1}{2}\big)^2-\big|\sigma_{a_j}\big|^2+2\Big(\overline{a_1^\dagger a_2}\cdot\overline{a_1a_2^\dagger}-\big|\overline{a_1a_2}\big|^2\Big),
\end{equation}
    where $\sigma_{N_j}\equiv\braket{\hat{a}_j^\dagger\hat{a}_j}-\big|\braket{\hat{a}_j}\big|^2$, and $\sigma_{a_j}\equiv\braket{\hat{a}_j^2}-\braket{\hat{a}_j}^2$. Taking the expectation values for the initial vacuum state $\ket{0,0}$ and by using \eqref{eq_tj} we get
\begin{equation}
    \begin{split}
        \sigma_{N_1}&=t_{12}t_{21}+t_{14}t_{23},\qquad
        \sigma_{a_1}=t_{11}t_{12}+t_{13}t_{14},\\
        \sigma_{N_2}&=t_{32}t_{41}+t_{34}t_{43},\qquad
        \sigma_{a_2}=t_{31}t_{32}+t_{33}t_{34}.
    \end{split}
\end{equation}
The rest of the terms are defined as $\overline{a_j a_i}=\frac{1}{2}\braket{\hat{a}_j\hat{a}_i+\hat{a}_i\hat{a}_j}-\braket{\hat{a}_j}\braket{\hat{a}_i}$. Taking the expectation values for the initial vacuum state we obtain
\begin{equation}
\begin{split}
    \overline{a_1^\dagger a_2}&=\left(t_{22}t_{31}+t_{21}t_{32}+t_{24}t_{33}+t_{23}t_{34}\right)/2,\\
    \overline{a_2^\dagger a_1}&=\left(t_{12}t_{41}+t_{11}t_{42}+t_{14}t_{43}+t_{13}t_{44}\right)/2,\\
    \overline{a_1 a_2}&=\left(t_{12}t_{31}+t_{11}t_{32}+t_{14}t_{33}+t_{13}t_{34}\right)/2.
\end{split}
\end{equation}
Here it is important to note that although the coefficients $t_{ij}$ change in time, the quantum universal invariant will remain constant during the evolution, independently of the specific coefficients forming the corresponding quadratic Hamiltonian~\cite{Dodonov_2000}. In Fig.~\ref{Fig_2}-(c) we confirm that $\tilde{\mathcal{D}}_{2}^{(2)}$ in Eq.~(\ref{ui}) is constant by integrating Eq.~(\ref{coeff_tjk}). Furthermore, it is easy to show that $\tilde{\mathcal{D}}_{2}^{(2)}=1/2$ at $t=0$ for the vacuum state $|0,0\rangle$. Therefore, this analysis proves that the $\alpha_j(t)$ functions, and consequently our time evolution operator $\hat{U}(t)$, are all correct.
\section{Conclusions}\label{conclusions}
We construct the exact time evolution operator of a non-stationary electromagnetic cavity having two coupled modes. Taking advantage of the fact that the set of operators in the corresponding effective Hamiltonian is closed under commutation we use the Wei-Norman theorem, and despite the fact that the Hamiltonian is time dependent,
the evolution operator can be written as a product of exponentials. In general, the method can be used for any given trajectory of one of the mirrors $q(t)$, which in turn defines the instantaneous eigenfrequency $\omega_k(t)$. On the other hand, we show how the intermode interaction has an effect on the generation of vacuum photons. However, we have used a particular form for $q(t)$, and we consider that an exploration of different trajectories could establish a better understanding of the intermode effect. Finally, due to the algebraic structure of the set of operators for the case of $N>3$ modes given by \eqref{eq:heff}, we expect that a similar procedure could lead to the exact solution of the problem.

\appendix
\section{}\label{Apen_A}
Here we show that the set of operators appearing in the effective Hamiltonian \eqref{eq:heff} is closed under commutation.  The set of operators for a single mode $\{ \hat a_i^{\dagger}\hat a_i, a_i^{\dagger 2}, \hat a_i^2\}$ is closed under commutation and  the commutation relations for the operators that couple the two modes  satisfy the commutation relations:
\begin{center}
\begin{table}[ht]
\begin{tabular}{|c||c|c|c|c|}
\hline
  &$\hat a_1^{\dagger}\hat a_2^{\dagger}$&$\hat a_1^{\dagger}\hat a_2$&$\hat a_1\hat a_2^{\dagger}$&$\hat a_1 \hat a_2$\\
 \hline
 \hline
 $\hat a_1^{\dagger}\hat a_2^{\dagger}$&0&$-\hat a_1^{\dagger 2}$&$-\hat a_2^{\dagger 2}$&$-(\hat n_1+\hat n_2+1)$\\
 \hline
 $\hat a_1^{\dagger}\hat a_2$&$\hat a_1^{\dagger 2}$&0&$\hat n_1-\hat n_2$&$-\hat a_2^{2}$\\
 \hline
 $\hat a_1 \hat a_2^{\dagger}$&$\hat a_2^{\dagger 2}$&$-\hat n_1+\hat n_2$&0&-$\hat a_1^2$\\
 \hline
 $\hat a_1\hat a_2$&$\hat n_1+\hat n_2+1$&$\hat a_2^{2}$&$\hat a_1^{2}$&0\\
 \hline
 \end{tabular}
 \caption{Commutation relations.}
\end{table}
\end{center}
The rest of the commutators that appear as we use the BCH relationships are given in the following tables:
 \begin{center}
 \begin{table}[ht]
\begin{tabular}{|c||c|c|c|c|c|c|}
\hline
 &$\hat a_1^{\dagger 2}$&$\hat a_2^{\dagger 2}$&$\hat n_1$&$\hat n_2$&$\hat a_1^2$&$\hat a_2^2$ \\
 \hline
 \hline
 $\hat a_1^{\dagger}\hat a_2^{\dagger}$&0&0&$-\hat a_1^{\dagger}\hat a_2^{\dagger}$&$-\hat a_1^{\dagger}\hat a_2^{\dagger}$&$-2\hat a_1\hat a_2^{\dagger}$&$-2\hat a_1^{\dagger}\hat a_2$ \\
 \hline
 $\hat a_1^{\dagger}\hat a_2$&0&$2\hat a_1^{\dagger}\hat a_2^{\dagger}$&$-\hat a_1^{\dagger}\hat a_2$&$\hat a_1^{\dagger}\hat a_2$&$-2\hat a_1 \hat a_2$&0\\
 \hline
 $\hat a_1\hat a_2^{\dagger}$&$2\hat a_1^{\dagger}\hat a_2^{\dagger}$&0&$\hat a_1 \hat a_2^{\dagger}$&$-\hat a_1\hat a_2^{\dagger}$&0&$-2\hat a_1 \hat a_2$ \\
 \hline
 $\hat a_1 \hat a_2$&$2\hat a_1^{\dagger}\hat a_2$&$2\hat a_1 \hat a_2^{\dagger}$&$\hat a_1 \hat a_2$&$\hat a_1 \hat a_2$&0&0\\
 \hline
 \end{tabular}
 \caption{Commutation relations.}
 \end{table}
 \end{center}
and
\begin{center}
\begin{table}[ht]
\begin{tabular}{|c||c|c|c|c|c|c|}
\hline
  &$\hat a_1^{\dagger 2}$&$\hat a_2^{\dagger 2}$&$\hat n_1$&$\hat n_2$&$\hat a_1^2$&$\hat a_2^2$ \\
 \hline
 \hline
$\hat a_1^{\dagger 2}$&0&0&$-2\hat a_1^{\dagger 2}$&0&$-4\hat n_1-2$&0 \\
\hline
$\hat a_2^{\dagger 2}$&0&0&0&$-2\hat a_2^{\dagger 2}$&0&$-4\hat n_2-2$\\
\hline
$\hat n_1$&$2\hat a_1^{\dagger 2}$&0&0&0&$-2\hat a_1^2$&0 \\
\hline
$\hat n_2$&0&$2\hat a_2^{\dagger 2}$&0&0&0&-$2\hat a_2^2$ \\
\hline
$\hat a_1^2$&$4\hat n_1+2$&0&$2\hat a_1^2$&0&0&0 \\
\hline
$\hat a_2^2$&0&$4\hat n_2+2$&0&$2\hat a_2^2$&0&0\\
 \hline
 \end{tabular}
 \caption{Commutation relations.}
 \end{table}
\end{center}

 Then, we see that the set of operators $\{ \hat a_1^{\dagger 2}, \hat a_2^{\dagger 2}, \hat a_1^{\dagger}\hat a_2^{\dagger}, \hat a_1^{\dagger}\hat a_2, \hat a_1\hat a_2^{\dagger}, \hat a_1^{\dagger}\hat a_1, \hat a_2^{\dagger}\hat a_2, \hat a_1^2, \hat a_2^2, \hat a_1\hat a_2,\hat1 \}$ is closed under commutation.

\section{}\label{Apen_B}
Here, taking advantage of the commutation relations shown above, it is more or less easy to obtain the following set of coupled, nonlinear, ordinary differential equations that we solved with Mathematica.
\begin{subequations}
\begin{equation}
\begin{split}
\dot{\alpha}_1&-\alpha _3 \dot{\alpha} _4+\left(\alpha _3 \alpha _4^2-2 \alpha _1 \alpha _4\right) \dot{\alpha} _5\\&+\left(\alpha _3 \alpha _5 \alpha _4^2+\alpha _3 \alpha _4-2 \alpha _1 \alpha _5 \alpha _4-2\alpha _1\right)\dot{\alpha} _6\\&+\left(-\alpha _3 \alpha _5 \alpha _4^2-\alpha _3 \alpha _4+2 \alpha _1 \alpha _5 \alpha _4\right)\dot{\alpha} _7\\
&+\left(4 e^{-2 \alpha _6} \alpha _1^2-4 e^{-2 \alpha_6} \alpha _3 \alpha _4 \alpha _1+e^{-2 \alpha _6} \alpha _3^2 \alpha _4^2\right)\dot{\alpha} _8\\
&+\big(e^{-2 \alpha _7} \alpha _3^2+e^{-2 \alpha _7} \alpha _4^2 \alpha _5^2 \alpha _3^2+2 e^{-2\alpha _7} \alpha _4 \alpha _5 \alpha _3^2\\&-4 e^{-2 \alpha _7} \alpha _1 \alpha _4 \alpha _5^2 \alpha _3-4 e^{-2 \alpha _7} \alpha _1 \alpha _5 \alpha _3\\&+4 e^{-2 \alpha _7} \alpha _1^2 \alpha_5^2\big)\dot{\alpha} _9+\big(-4 e^{-\alpha _6-\alpha _7} \alpha _5 \alpha _1^2\\&+2 e^{-\alpha _6-\alpha _7} \alpha _3 \alpha _1\\&+4 e^{-\alpha _6-\alpha _7} \alpha _3 \alpha _4 \alpha _5 \alpha_1\\&-e^{-\alpha _6-\alpha _7} \alpha _3^2 \alpha _4-e^{-\alpha _6-\alpha _7} \alpha _3^2 \alpha _4^2 \alpha _5\big)\dot{\alpha} _{10}+\mathrm{i}f_1=0,
\end{split}
\end{equation}
\begin{equation}
\begin{split}
\dot{\alpha}_2&+\left(2 \alpha _2 \alpha _4-\alpha _3\right)\dot{\alpha}_5+\left(2 \alpha _2 \alpha _4 \alpha _5-\alpha _3 \alpha _5\right)\dot{\alpha}_6\\&+\left(-2 \alpha _4 \alpha _5 \alpha _2-2 \alpha _2+\alpha _3 \alpha _5\right)\dot{\alpha} _7\\
&+\left(e^{-2 \alpha _6} \alpha _3^2-4 e^{-2\alpha _6} \alpha _2 \alpha _4 \alpha _3+4 e^{-2 \alpha _6} \alpha _2^2 \alpha _4^2\right)\dot{\alpha} _8\\
&+\big(4 e^{-2 \alpha _7} \alpha _2^2+4 e^{-2 \alpha _7} \alpha _4^2 \alpha _5^2 \alpha _2^2+8 e^{-2 \alpha _7} \alpha _4 \alpha _5 \alpha _2^2\\&-4 e^{-2 \alpha _7} \alpha _3\alpha _4 \alpha _5^2 \alpha _2-4 e^{-2 \alpha _7} \alpha _3 \alpha _5 \alpha _2+e^{-2 \alpha _7} \alpha _3^2 \alpha _5^2\big)\dot{\alpha} _9\\
&+\big(-4 e^{-\alpha _6-\alpha _7} \alpha _4 \alpha _2^2-4 e^{-\alpha _6-\alpha _7} \alpha _4^2 \alpha _5 \alpha _2^2\\&+2 e^{-\alpha_6-\alpha _7} \alpha _3 \alpha _2\\&+4 e^{-\alpha _6-\alpha _7} \alpha _3 \alpha _4 \alpha _5 \alpha _2-e^{-\alpha _6-\alpha _7} \alpha _3^2 \alpha _5\big)\dot{\alpha} _{10}+\mathrm{i}f_2=0,
\end{split}
\end{equation}
\begin{equation}
\begin{split}
\dot{\alpha} _3&-2\alpha _2\dot{\alpha}_4+\left(2 \alpha _2 \alpha _4^2-2 \alpha _1\right)\dot{\alpha }_5\\&+\left(2 \alpha _2 \alpha _5 \alpha _4^2+2 \alpha _2 \alpha _4-\alpha _3-2 \alpha _1 \alpha _5\right)\dot{\alpha}_6\\
&+\left(-2 \alpha _2 \alpha _5 \alpha _4^2-2 \alpha _2 \alpha _4-\alpha _3+2\alpha _1 \alpha _5\right)\dot{\alpha} _7\\
&+\big(-2 e^{-2 \alpha _6} \alpha _4 \alpha _3^2+4 e^{-2 \alpha _6} \alpha _2 \alpha _4^2 \alpha _3\\&+4 e^{-2 \alpha _6} \alpha _1 \alpha _3-8 e^{-2 \alpha _6} \alpha _1 \alpha _2 \alpha _4\big)\dot{\alpha} _8\\
&+\big(-2 e^{-2 \alpha _7} \alpha_4 \alpha _5^2 \alpha _3^2-2 e^{-2 \alpha _7} \alpha _5 \alpha _3^2\\&+4 e^{-2 \alpha _7} \alpha _2 \alpha _4^2 \alpha _5^2 \alpha _3+4 e^{-2 \alpha _7} \alpha _1 \alpha _5^2 \alpha _3\\&+4 e^{-2 \alpha _7} \alpha _2 \alpha _3+8 e^{-2 \alpha _7} \alpha _2 \alpha _4 \alpha _5 \alpha_3\\&-8 e^{-2 \alpha _7} \alpha _1 \alpha _2 \alpha _4 \alpha _5^2-8 e^{-2 \alpha _7} \alpha _1 \alpha _2 \alpha _5\big)\dot{\alpha}_9\\
&+\big(e^{-\alpha _6-\alpha _7} \alpha _3^2+2 e^{-\alpha _6-\alpha _7} \alpha _4 \alpha _5 \alpha _3^2\\&-4 e^{-\alpha _6-\alpha _7} \alpha _2 \alpha_4 \alpha _3-4 e^{-\alpha _6-\alpha _7} \alpha _2 \alpha _4^2 \alpha _5 \alpha _3\\&-4 e^{-\alpha _6-\alpha _7} \alpha _1 \alpha _5 \alpha _3+4 e^{-\alpha _6-\alpha _7} \alpha _1 \alpha _2\\&+8 e^{-\alpha _6-\alpha _7} \alpha _1 \alpha _2 \alpha _4 \alpha _5\big)\dot{\alpha}_{10}+\mathrm{i}f_3=0,
\end{split}
\end{equation}
\begin{equation}
\begin{split}
\dot{\alpha} _4&-\alpha _4^2\dot{\alpha} _5-\left(\alpha _5 \alpha _4^2+\alpha _4\right)\dot{\alpha} _6+\left(\alpha _5 \alpha _4^2+\alpha _4\right)\dot{\alpha} _7\\&+\left(4 e^{-2 \alpha _6} \alpha _1 \alpha _4-2 e^{-2 \alpha _6} \alpha _3 \alpha _4^2\right)\dot{\alpha} _8\\
&+\big(-2 e^{-2\alpha _7} \alpha _3 \alpha _4^2 \alpha _5^2+4 e^{-2 \alpha _7} \alpha _1 \alpha _4 \alpha _5^2\\&+4 e^{-2 \alpha _7} \alpha _1 \alpha _5-4 e^{-2 \alpha _7} \alpha _3 \alpha _4 \alpha _5-2 e^{-2 \alpha _7} \alpha _3\big)\dot{\alpha} _9\\
&+\big(2 e^{-\alpha _6-\alpha _7} \alpha _3\alpha _5 \alpha _4^2+2 e^{-\alpha _6-\alpha _7} \alpha _3 \alpha _4\\&-4 e^{-\alpha _6-\alpha _7} \alpha _1 \alpha _5 \alpha _4-2 e^{-\alpha _6-\alpha _7} \alpha _1\big)\dot{\alpha} _{10}+\mathrm{i}f_4=0,
\end{split}
\end{equation}
\begin{equation}
\begin{split}
\dot{\alpha} _5&+\alpha _5\dot{\alpha}_6-\alpha _5\dot{\alpha} _7+\left(4 e^{-2 \alpha _6} \alpha _2 \alpha _4-2 e^{-2 \alpha _6} \alpha _3\right)\dot{\alpha} _8\\&+\left(-2 e^{-2 \alpha _7} \alpha _3 \alpha _5^2+4 e^{-2 \alpha _7} \alpha _2 \alpha _4 \alpha _5^2+4 e^{-2 \alpha _7} \alpha _2 \alpha_5\right)\dot{\alpha}_9\\&+\big(-2 e^{-\alpha _6-\alpha _7} \alpha _2-4 e^{-\alpha _6-\alpha _7} \alpha _4 \alpha _5 \alpha _2\\&+2 e^{-\alpha _6-\alpha _7} \alpha _3 \alpha_5\big)\dot{\alpha}_{10}+\mathrm{i}f_5=0,\\
\end{split}
\end{equation}
\begin{equation}
\begin{split}
\alpha _4\dot{\alpha}_5&+\left(\alpha _4 \alpha _5+1\right)\dot{\alpha}_6-\alpha _4 \alpha _5\dot{\alpha}_7\\&+\left(2 e^{-2 \alpha _6} \alpha _3 \alpha _4-4 e^{-2 \alpha _6} \alpha _1\right)\dot{\alpha} _8\\
&+\bigg(-4 e^{-2 \alpha _7} \alpha _1 \alpha _5^2+2 e^{-2 \alpha _7} \alpha _3 \alpha _4\alpha _5^2\\&+2 e^{-2 \alpha _7} \alpha _3 \alpha _5\bigg)\dot{\alpha}_9\\
&+\big(-e^{-\alpha _6-\alpha _7} \alpha _3-2 e^{-\alpha _6-\alpha _7} \alpha _4 \alpha _5 \alpha _3\\&+4 e^{-\alpha _6-\alpha _7} \alpha _1 \alpha _5\big)\dot{\alpha}_{10}+\mathrm{i}f_6=0,
\end{split}
\end{equation}
\begin{equation}
\begin{split}
-\alpha _4\dot{\alpha}_5&-\alpha _4 \alpha _5\dot{\alpha} _6+\left(\alpha _4 \alpha _5+1\right)\dot{\alpha}_7\\&+\left(2 e^{-2 \alpha _6} \alpha _3 \alpha _4-4 e^{-2 \alpha _6} \alpha _2 \alpha _4^2\right)\dot{\alpha}_8\\
&+\big(-4 e^{-2 \alpha _7} \alpha _2 \alpha _4^2 \alpha _5^2+2 e^{-2 \alpha_7} \alpha _3 \alpha _4 \alpha _5^2\\&+2 e^{-2 \alpha _7} \alpha _3 \alpha _5-8 e^{-2 \alpha _7} \alpha _2 \alpha _4 \alpha _5-4 e^{-2 \alpha _7} \alpha _2\big)\dot{\alpha} _9\\
&+\big(4 e^{-\alpha _6-\alpha _7} \alpha _2 \alpha _5 \alpha _4^2+4 e^{-\alpha _6-\alpha _7} \alpha _2\alpha _4\\&-2 e^{-\alpha _6-\alpha _7} \alpha _3 \alpha _5 \alpha _4-e^{-\alpha _6-\alpha _7} \alpha _3\big)\dot{\alpha} _{10}+\mathrm{i}f_7=0,
\end{split}
\end{equation}
\begin{equation}
\begin{split}
e^{-2 \alpha _6}\dot{\alpha}_8+e^{-2 \alpha _7}\alpha _5^2\dot{\alpha} _9-e^{-\alpha _6-\alpha _7}\alpha _5\dot{\alpha}_{10}+\mathrm{i}f_8=0,\\
\end{split}
\end{equation}
\begin{equation}
\begin{split}
&e^{-2 \alpha _6}\alpha _4^2\dot{\alpha}_8+\left(e^{-2 \alpha _7} \alpha _4^2 \alpha _5^2+2 e^{-2 \alpha _7} \alpha _4 \alpha _5+e^{-2 \alpha _7}\right)\dot{\alpha}_9\\&-\left(e^{-\alpha _6-\alpha _7} \alpha _5 \alpha _4^2+e^{-\alpha _6-\alpha _7} \alpha _4\right)\dot{\alpha}_{10}+\mathrm{i}f_9=0,\\
\end{split}
\end{equation}
\begin{equation}
\begin{split}
&-2 e^{-2 \alpha _6}\alpha _4\dot{\alpha}_8+\left(-2 e^{-2 \alpha _7} \alpha _4 \alpha _5^2-2 e^{-2 \alpha _7} \alpha _5\right)\dot{\alpha} _9\\&+\left(2 e^{-\alpha _6-\alpha _7} \alpha _4 \alpha _5+e^{-\alpha _6-\alpha _7}\right)\dot{\alpha}_{10}+\mathrm{i}f_{10}=0,
\end{split}
\end{equation}
\begin{equation}
\begin{split}
&\left(-2 e^{-2 \alpha _6} \alpha _2 \alpha _4^2+2 e^{-2 \alpha _6} \alpha _3 \alpha _4-2 e^{-2 \alpha _6} \alpha _1\right)\dot{\alpha}_8\\&+\big(-2 e^{-2 \alpha _7} \alpha _2 \alpha _4^2 \alpha _5^2-2 e^{-2 \alpha _7} \alpha _1 \alpha _5^2+2 e^{-2 \alpha _7} \alpha _3 \alpha _4 \alpha_5^2\\&+2 e^{-2 \alpha _7} \alpha _3 \alpha _5-4 e^{-2 \alpha _7} \alpha _2 \alpha _4 \alpha _5-2 e^{-2 \alpha _7} \alpha _2\big)\dot{\alpha}_9\\
&+\big(2 e^{-\alpha _6-\alpha _7} \alpha _2 \alpha _5 \alpha _4^2+2 e^{-\alpha _6-\alpha _7} \alpha _2 \alpha _4\\&-2 e^{-\alpha _6-\alpha_7} \alpha _3 \alpha _5 \alpha _4-e^{-\alpha _6-\alpha _7} \alpha _3+2 e^{-\alpha _6-\alpha _7} \alpha _1 \alpha_5\big)\dot{\alpha}_{10}\\&+\dot{\alpha}_{11}+\mathrm{i}f_{11}=0.
\end{split}
\end{equation}
\end{subequations}
Consequently, we can identify $M_{1j}$, and finally we get $M_ {nj}$. Therefore it is possible to write the set of differential equations in compact form as:
\begin{equation}\label{Eq:alphasM}
\bm{M}\dot{\bm{\alpha}}(t)+\mathrm{i}\bm{f}(t)=0.
\end{equation}
Hence, the system of differential equations given by \eqref{alphas} is completely defined and in principle can be solved for any set of functions $f_j (t)$.

Finally, we need the expressions of the coefficients $t_{ij}$ that come from applying the evolution operator $\hat{U}(t)$ to the creation and annihilation operators of each mode, $\hat{a}_j^\dagger(t)=\hat{U}^\dagger(t)\hat{a}_j^\dagger\hat{U}(t)$ and $\hat{a}_j(t)=\hat{U}^\dagger\hat{a}_j\hat{U}(t)$. These coefficients are:
\begin{equation}\label{coeff_tjk}
\begin{split}
t_{11}=&e^{\alpha_6}(1+\alpha_4\alpha_5)-2e^{-\alpha_6}(2\alpha_1-\alpha_3\alpha_4)\alpha_8\\&-e^{-\alpha_7}(\alpha_3-2\alpha_1\alpha_5+\alpha_3\alpha_4\alpha_5)\alpha_{10}, \\
t_{12}=&e^{-\alpha_6}(2\alpha_1-\alpha_3\alpha_4), \\
t_{13}=&e^{\alpha_7}\alpha_4-e^{-\alpha_7}(\alpha_3-4\alpha_1\alpha_5+2\alpha_3\alpha_4\alpha_5)\alpha_9\\&-e^{-\alpha_6}(2\alpha_1-\alpha_3\alpha_4)\alpha_{10}, \\
t_{14}=&e^{-\alpha_7}(\alpha_3-2\alpha_1\alpha_5+\alpha_3\alpha_4\alpha_5), \\
t_{21}=&-2\alpha_8e^{-\alpha_6}+\alpha_5\alpha_{10}e^{-\alpha_7}, \\
t_{22}=&e^{-\alpha_6}, \\
t_{23}=& 2e^{-\alpha_7}\alpha_5\alpha_9-e^{-\alpha_6}\alpha_{10}, \\
t_{24}=& -e^{-\alpha_7}\alpha_5, \\
t_{31}=& e^{\alpha_6}\alpha_5-2e^{-\alpha_6}(\alpha_3-2\alpha_2\alpha_4)\alpha_8\\&-e^{-\alpha_7}(2\alpha_2-\alpha_3\alpha_5+2\alpha_2\alpha_4\alpha_5)\alpha_{10}, \\
t_{32}=&e^{-\alpha_6}(\alpha_3-2\alpha_2\alpha_4), \\
t_{33}=&e^{\alpha_7}-2e^{-\alpha_7}(2\alpha_2-\alpha_3\alpha_5+2\alpha_2\alpha_4\alpha_5)\alpha_9\\&-e^{-\alpha_6}(\alpha_3-2\alpha_2\alpha_4) \alpha_{10},\\
t_{34}=&e^{-\alpha_7}(2\alpha_2-\alpha_3\alpha_5+2\alpha_2\alpha_4\alpha_5), \\
t_{41}=&2e^{-\alpha_6}\alpha_4\alpha_8-e^{-\alpha_7}(1+\alpha_4\alpha_5)\alpha_{10}, \\
t_{42}=&-e^{-\alpha_6}\alpha_4, \\
t_{43}=&-2e^{-\alpha_7}(1+\alpha_4\alpha_5)\alpha_9+e^{-\alpha_6}\alpha_4\alpha_{10}, \\
t_{44}=&e^{-\alpha_7}(1+\alpha_4\alpha_5).
\end{split}
\end{equation}
\section*{Acknowledgments}J.R. and I.R.P. thank Reyes Garc\'{\i}a for the maintenance of our computers, and acknowledge partial support from Direcci\'on General de Asuntos del Personal Acad\'emico, Universidad Nacional Aut\'onoma de M\'exico (DGAPA UNAM) through project PAPIIT IN 1111119.
%

\end{document}